\pdfoutput=1
\documentclass[
  ,draft            % use draft while you are working on the paper
  ]
  {aipproc}

\layoutstyle{8x11single}

\include{command}

\begin{document}

\title{Trigger, Reconstruction and Physics Performances in LHCb}



\author{C.Lazzeroni. For the LHCb Collaboration}{
  address={Cavendish Laboratory, JJ Thomson Avenue,
 Cambridge,
 CB3 0HE,
 United Kingdom}
}

\begin{abstract}
The LHCb experiment at LHC is a single-armed spectrometer
designed to pursue extensive, high precision studies
of CP violation and rare phenomena in $b$ hadron decays.
In this contribution, the trigger and reconstruction performance
are summarised, and the expected performance for selected
$b$ physics measurements is discussed.
\end{abstract}

\maketitle

%%%%%%%%%%%%%%%%%%%%%%%%%%%%%%%%%%%%%%%%%%%%
%% MAINMATTER
%%%%%%%%%%%%%%%%%%%%%%%%%%%%%%%%%%%%%%%%%%%%

\section{Introduction}

LHCb is one of the four major experiments that will take data at the LHC, 
due to start operation in 2007.
The primary aims of LHCb are to perform precision tests 
of CP violation and to search for new physics in $b$ hadron decays.
About $10^{12}$ $b \bar{b}$ 
pairs will be produced in LHCb per $10^7$ seconds (a nominal year)
in $pp$ collisions
with a luminosity of $2 \times 10^{32} \ cm^{-2} s^{-1}$. 
A large, high-purity 
sample of $b$ hadrons, decaying in
a variety of channels, will be accumulated. 
%Together with the excellent knowledge already obtained on the angle $\beta$ 
%from experiments at $e^+e^-$ colliders,
%and the information available on the sides of the triangle, a 
%measurement of the other two angles will over-constrain 
%the unitary triangle 
%and may allow detection of NP contributions to CP violation.
%The LHCb physics programme will focus on high precision studies of 
%CP violation and rare phenomena in b hadron decays.
%At the LHCb interaction point $10^{12}$ b hadron events per $10^7$
%second (a nominal year) will be produced in $pp$ collisions
%with a luminosity of $2 \times 10^{32} \ cm^{-2} s^{-1}$.
LHCb will perform a detailed study of B meson mixing,
precise measurements of the angles
of the unitary triangle and investigations of 
rare decays in $b$ hadrons, looking for new physics 
in loop-induced processes.

The LHCb detector is optimised to reach these physics goals.
Here a brief description of the reconstruction performance is given,
and the sensitivities in typical channels for the study of $B_s$ mixing,
CP violation and rare decays are summarised.

\section{Trigger and Event Reconstruction performance}

At the LHC energies, the production of $b$ and $\bar{b}$ quarks
is predominantly along the beam axis, and
highly correlated with each other, so that if
one $b$ goes forward into the detector acceptance, the corresponding $\bar{b}$
products are also captured with high probability.
This leads to high effienciency for a single arm spectrometer design, 
like LHCb, where
only a factor $\sim$2 is lost compared to a twin-arm spectrometer.
%$4\pi$ geometry.
Moreover, owing to an extended acceptance in the forward direction 
(higher $\eta$),
a factor 2 better $b\bar{b}$ cross section is expected at LHCb
compared to the ATLAS and CMS experiments.
At the LHCb luminosity, events are dominated by single proton-proton
intereactions and the occupancy in the detector remains low.

The LHCb detector comprises a beam pipe, a Vertex Locator,
a tracking system with a dipole magnet, two Ring Imaging
Cherenkov detectors, an electromagnetic and a hadronic calorimeter
and a muon system. 
To achieve its physics goals, efficient trigger, 
exclusive signal reconstruction, good
proper time resolution and flavour tagging are essential for LHCb.
For a detailed description of the detector,
see \cite{VeloTDR,ITTDR,OTTDR,RichTDR,CaloTDR,MuonTDR}.
For an update on the installation and commisioning status, see \cite{Bolek}.

\subsection{Trigger}

The experiment will use various levels of
trigger to reduce
the 10~MHz rate of visible interactions to the 2~kHz
that will be stored.
The $b$-quark production cross section of $\sim 0.5$ mb at 
$\sqrt s \sim $14 TeV $pp$ interactions
is only a small fraction of the total visible cross section of $\sim 100$ mb.
The LHCb trigger fully exploits the $b$-decay topology, characterised by 
significant transverse momentum
due to the high $b$-quark mass, and long lifetime yielding large
impact parameter (IP) values.
The first level trigger (Level 0, hardware) relies on high-$p_T$ $e,\gamma , 
\pi^0, \mu , h$ candidates, decreasing
the rate to 1~MHz. The higher level trigger (HLT) is software based and 
makes use of the full detector information. Four independent trigger
streams are defined, depending on the Level 0 output:
$\mu$, $\mu + h$, $h$, and $ecal$ streams.
The $p_T$ and IP discriminations
are then used to reduce the output to 2~kHz. 
The overall trigger efficiency varies
from 30\% to 80\% depending on the signal channels.

\subsection{Reconstruction}

%To achieve its physics goals, exclusive signal reconstruction, good
%proper time resolution and flavour tagging are essential for LHCb.

Exclusive decay reconstruction requires good mass resolution, which is linked
to excellent momentum resolution. 
For tracks with a momentum higher than 10 GeV/c the average efficiency
is $\sim$95\%, tuned to have a ghost rate of $\sim$4\% for $p_t >
0.5$ GeV/c \cite{Jonesc}.
The average track momentum resolution is $\delta p/p = 0.37$\%. 

\begin{figure}[!t]
  \includegraphics[height=.24\textheight]{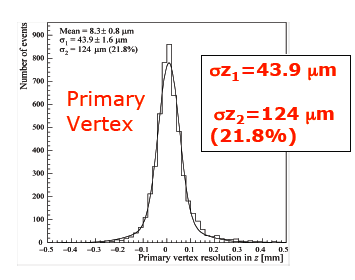}
  \hspace{3.5mm}
  \includegraphics[height=.24\textheight]{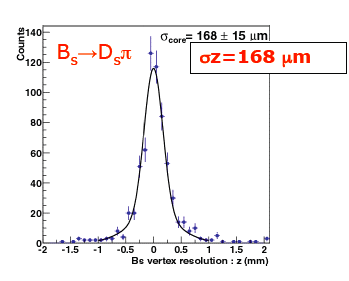}
  \hspace{3.5mm}
  \includegraphics[height=.22\textheight]{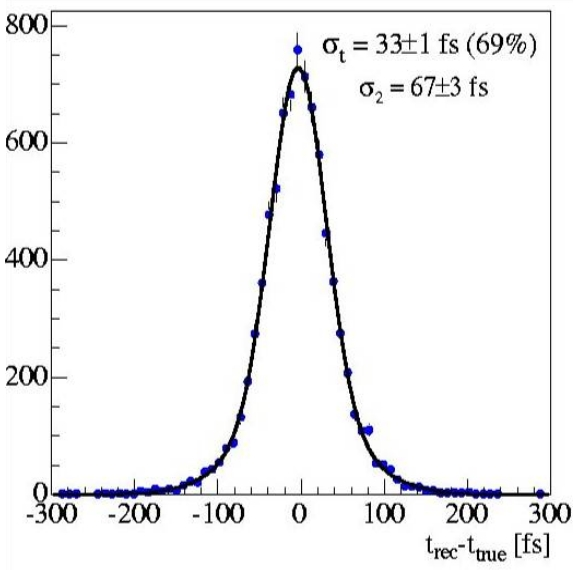}
  \caption{Primary vertex resolution (left), secondary vertex 
resolution (centre), proper time resolution (right).}
  \label{vtx}
\end{figure}

The proper time of a $b$ hadron decay is determined from the distance
between its production and decay vertex and from its momentum.
Using a double Gaussian fit to the distribution of the vertex residuals,
the resolution on the primary vertex position along
the beam direction (z coordinate) is measured to be 44 $\mu$m,
with 22\% of the events being in the second Gaussian which is 124 $\mu$m
wide (see Fig. \ref{vtx} (left)). 
A resolution of 168 $\mu$m is obtained for the $B_s \rightarrow D_s K$
decay vertex, taken as an example,
using a simple Gaussian fit (see Fig. \ref{vtx} (centre)).
Therefore, the proper time resolution is usually 
dominated by the resolution on the
B decay vertex. 
The proper time resolution for the same decay is shown 
in Fig. \ref{vtx} (right). 
Using a double Gaussian fit the core resolution is measured to be 33 $fs$.
The second Gaussian accounts for 31\% of the events and has a width of 67 $fs$.

Efficient particle identification in the momentum range 2-100 GeV/c is
needed for flavour tagging and background rejection. The average 
kaon identification efficiency for momenta between 2 and 100 GeV/c
is 83\% with a $\pi$ mis-identification rate of 6\%. Different optimisations
can be performed, in order to improve the mis-identification rate 
at the expences of the efficiency. 
A detailed review of tracking performance and particle identification
can be found in \cite{Jonesc}.

\begin{figure}[!t]
  \includegraphics[height=.24\textheight]{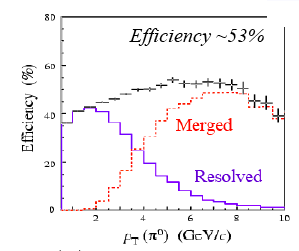}
  \hspace{5mm}
  \includegraphics[height=.24\textheight]{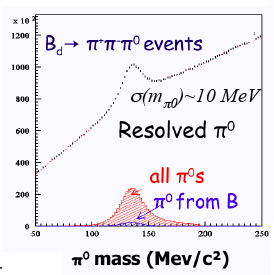}
  \hspace{5mm}
  \includegraphics[height=.24\textheight]{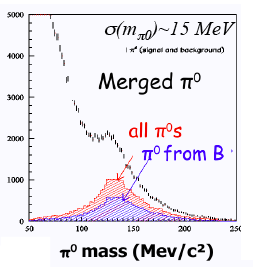}
  \caption{Efficiency for $\pi^0$ reconstruction as a function
of $p_T$ (left), $\pi^0$ mass resolution for resolved $\pi^0$s (centre),
$\pi^0$ mass resolution for merged $\pi^0$s (right).}
  \label{neutral}
\end{figure}

Neutral particle reconstruction is performed using both resolved 
(separate) clusters and merged (overlapping) cluster shapes
in the electromagnetic calorimeter.
A mean efficiency of $\sim$53\% is obtained for $B^0 \rightarrow \pi^+
\pi^- \pi^0$ events, as shown in Fig. \ref{neutral} (left) as a function
of the $\pi^0$ transverse momentum. The $\pi^0$ mass resolution 
is shown in Fig. \ref{neutral} (centre) and (right) for resolved and
merged $\pi^0$ respectively.

\begin{figure}[!t]
  \includegraphics[height=.24\textheight]{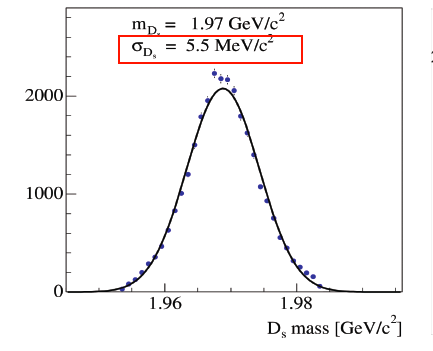}
  \hspace{10mm}
  \includegraphics[height=.24\textheight]{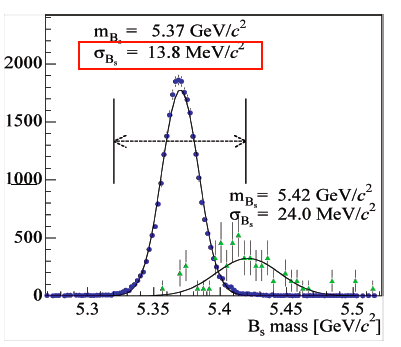}
%  \hspace{5mm}
%  \includegraphics[height=.24\textheight]{tagtable}
  \caption{$D_s$ mass resolution (left),
$B_S$ mass resolution (right). The arrow indicates the $3\sigma$ region.} 
%Tagging power (right).}
  \label{resolution}
\end{figure}

As an example of the mass resolutions that can be achieved,
the reconstructed $D_s$ and $B_s$ mass distributions for 
$B_s \rightarrow D_s(KK\pi) K$ events
are shown in Fig \ref{resolution} (left, centre). 
The mass resolutions are $\sim 5.5$ MeV/c$^2$ for $D_s$ and $\sim 14$
MeV/c$^2$ for $B_s$. The second peak in the $B_s$ distribution
corresponds to $B_s \rightarrow D_s(KK\pi) \pi$ where the $\pi$ has been
mis-identified as a $K$.

The identification of the initial b-quark charge (flavour)
of a reconstructed $b$ hadron decay is performed using opposite-side and
same-side tagging algorithms. To determine the flavour of the accompanying B,
opposite-side tagging uses the charge of:
a) leptons from a semileptonic decay; b) a kaon from a $b \rightarrow c 
\rightarrow s$ chain; c) the charge of all particles in a jet or at a vertex. 
Same-side tagging uses
the charge of fragmentation particles which are correlated in phase
space with the signal B meson to determine its flavour.
Figure \ref{table} (left) lists the tagging power ($\epsilon D^2 = 
\epsilon (1 - 2 w)^2$), where $\epsilon$ is the tagging efficiency
and $w$ is the wrong tagging fraction) of each tagging category
and the combined tagging power for $B_d$ and $B_s$ mesons. The range
of values depends on the signal channel considered.

For further details on reconstruction performances, see \cite{ReopTDR}.

%\begin{minipage}[b]{3.5cm}
%
%\begin{table}
%\begin{tabular}{rr}
%\hline
%\tablehead{1}{r}{b}{Tag} 
%& \tablehead{1}{r}{b}{Tagging power} \\
%\hline
%$\mu$ opposite side &  0.7\% - 1.8\% \\
%$e$ opposite side &  0.4\% - 0.6\% \\
%$K$ opposite side  &  1.6\%  -  2.4\% \\
%Jet/Vertex charge  &  0.9\%  -  1.3\% \\
%$\pi$ same side ($B^0$) & 0.8\% - 1.0\% \\
%$K$ same side ($B_s$) &  2.7\%  -  3.3\% \\
%\hline
%Combined ($B^0$) & 4\% - 5\% \\
%Combined ($B_s$) & 7\% - 9\% \\
%\hline
%\end{tabular}
%\caption{Tagging power}
%\label{tagging}
%\end{table}
%
%\end{minipage}
%
%\hfill
%
%\begin{minipage}[t]{5.5cm}
%
%\begin{table}
%\begin{tabular}{rrr}
%\hline
%\tablehead{1}{r}{b}{Mode(+cc)} 
%& \tablehead{1}{r}{b}{Yield} & \tablehead{1}{r}{b}{S/B(90\%CL)}\\
%\hline
%$B^0 \rightarrow D^0 (K^+ \pi^-) K^{*0}$ &  3.4k & $>2$ \\
%$B^0 \rightarrow D^0 (K^- \pi^+) K^{*0}$ &  0.5k & $>0.3$ \\
%$B^0 \rightarrow D^0_{CP} (K^+ K^-) K^{*0}$ &  0.6k & $>0.3$ \\
%\hline
%\end{tabular}
%\caption{Annual yield}
%\label{yield}
%\end{table}

%\end{minipage}

\section{Physics performance}

Selected baseline physics measurements are discussed below.
All event samples were produced using the full detector simulation, 
realistic digitization
and reconstruction with full pattern recognition and realistic
trigger simulation. Toy Monte Carlo algorithms were used for
sensitivity studies. The background was assumed to come from 
$b\bar{b}$ inclusive events, having similar topologies.
All sensitivities are relative to one year of data taking or $10^7$ s,
equivalent to $2 fb^{-1}$ accumulated statistics, unless
otherwise specified. For the estimates
of the background over signal ratio (B/S),
the 90\% CL is used throughout this paper.

\subsection{$B_s \bar{B_s}$ mixing}

The LHCb precision for measuring 
$B_s \bar{B_s}$ oscillations has been estimated using the decay
$B_s \rightarrow D_s \pi$.
A sample of 80k events ($B/S < 0.3$) 
will be collected per year of running, with a proper 
time resolution of $\sim 40 \ fs$.
A clear observation of the oscillation pattern can be performed, as seen
in Fig.\ref{sec12} (left). Given the value of 
$\Delta m_s$ recently measured \cite{CDF,D0},
LHCb will be able to measure it with much less than
$2 fb^{-1}$ of data, and a high precision measurement is
expected in 1 year:
$\sigma_{stat} (\Delta m_s)\sim 0.01 \ ps^{-1}$.

The $B_s \rightarrow J/\Psi \phi$ decay measures simultaneously
$\phi_s = -2 \chi$ and $\Delta \Gamma_s / \Gamma_s$,
the oscillation phase and the decay width difference between the
two CP eigenstates. The values
for those quantities
predicted by the Standard Model are respectively -0.04 and 0.1. 
This decay mode provides
high statistics ($\sim$120k events/year, $B/S < 0.33$), 
but requires partial wave analysis since it contains both CP-even
and CP-odd contributions.
The sensitivities after 1 year of running are $\sigma(\phi_s) \sim 0.03$
and $\sigma(\Delta \Gamma_s / \Gamma_s) \sim 0.02$; the first is comparable
to the Standard Model prediction, while the second is significantly better.
Using the CP modes $B \rightarrow J/\Psi \eta, \eta_c \phi$, 
the sensitivity is improved to $\sigma(\phi_s) \sim 0.013$ after 5 years.

\subsection{Measurement of $\alpha$ and $\gamma$}

LHCb will provide precise measurements of all the angles of the 
unitary triangle. 
Together with the excellent knowledge already obtained on the angle $\beta$ 
from experiments at $e^+e^-$ colliders,
and the information available on the sides of the triangle, a 
measurement of the other two angles will over-constrain 
the unitary triangle 
and may allow detection of NP contributions to CP violation. 

\begin{figure}[!t]
  \includegraphics[height=.24\textheight]{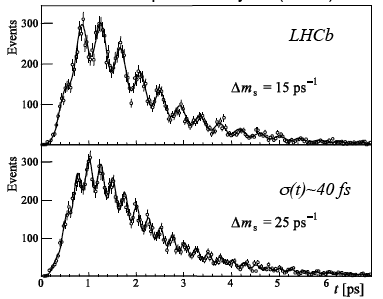}
  \hspace{10mm}
  \includegraphics[height=.24\textheight]{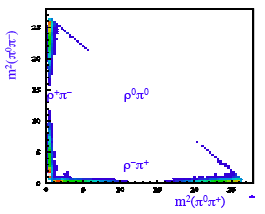}
  \caption{Distribution of $B_s \rightarrow D_s \pi$ proper time (ps)
after 1 year of data taking, for $\Delta m_s = 15, 25 \ ps^{-1}$ (left).
Dalitz plot for the decay $B \rightarrow \pi^0 \pi^+ \pi^-$ (right).}
  \label{sec12}
\end{figure}

The sensitivity for the angle $\alpha$ has been studied using the decay
$B_d \rightarrow \rho \pi$. 
A sample of 14k events will be reconstructed 
per year, with a selection based on multivariate analysis, 
and a ratio $B/S < 1$. Using a time-dependent Dalitz fit method
\cite{alpha}, the angle $\alpha$ can be determined with one year of data
to a precision of $10^\circ$ (see Fig.\ref{sec12} (right) for the Dalitz plot).

%\subsection{Measurement of $\gamma$}
The angle $\gamma$ can be measured at LHCb using several methods.

%\begin{itemize}
 
\begin{figure}[!t]
  \includegraphics[height=.24\textheight]{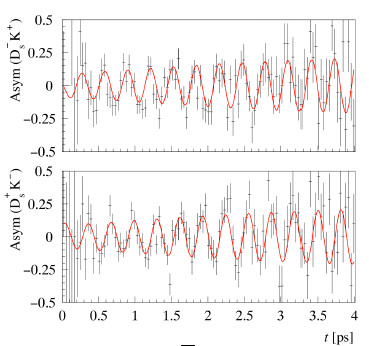}
  \hspace{10mm}
  \includegraphics[height=.24\textheight]{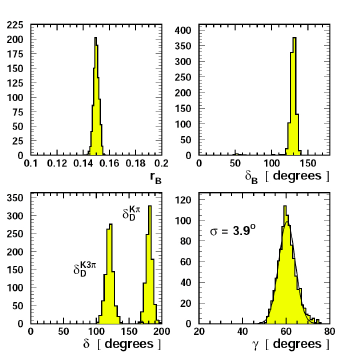}
  \caption{Distribution of $B_s \rightarrow D_s K$ asymmetry
versus proper time (ps) for $D_s^+ K^-$ and $D_s^- K^+$ (left).
Example of fit results for $r_B$, $\delta_B$, $\delta_D$ and $\gamma$
in $B^\pm \rightarrow D^0 K^\pm$ using the ADS method with no background.}
  \label{dsk}
\end{figure}

The decay $B_s \rightarrow D_s K$, where only tree diagrams contribute,
should provide a measurement of $\gamma -2 \chi$ free from new 
physics effects. The mixing phase $2 \chi$ will then be measured
separately, as illustrated above. Here the $K/\pi$ separation
significantly suppresses reflections from $B_s \rightarrow D_s \pi$.
An annual yield of 5.4k events is expected, with a ratio $B/S < 1$.
The measurement of $\gamma -2 \chi$ will be performed using the
time-dependent rates of $B_s \rightarrow D_s^+ K^-$ and 
$B_s \rightarrow D_s^- K^+$,
and their charge conjugates (see Fig \ref{dsk}).
For $\Delta m_s = 20 ps^{-1}$ this measurement gives a precision
$\sigma(\gamma)$ of 14$^\circ $ with 2$fb^{-1}$.

$\gamma$ can be extracted from the interference of 
$B^\pm \rightarrow D^0 K^\pm$
and $B^\pm \rightarrow \bar{D^0} K^\pm$ when $D^0$ and $\bar{D^0}$ decay
to a common final state. 
The decay amplitude can be parameterised as $A(B^- \rightarrow \bar{D^0} K^-)=
A_B r_B e^{i(\delta - \gamma)}$ where $A_B = A(B^- \rightarrow D^0 K^-)$,
$r_B$ is the relative colour and CKM suppression between the two modes, and
$\delta$ is the strong phase difference.
Two types of decays are currently under study
in LHCb: Cabibbo favoured self-conjugate decays (like $K_s \pi\pi$), the
sensitivity of which
is under study, and Cabibbo favoured/doubly Cabibbo suppressed
modes (like $K\pi$,$K\pi\pi\pi$). 
The latter study is based on the ADS method \cite{ads,ads1}:
the relative rates of $B^+ \rightarrow D K^+$
and $B^- \rightarrow D K^-$ depend on the parameters $\gamma$,
$r_B$, $\delta_B$, $r_D$, $\delta_D$. Taking $r_B=0.15$, $\sim$60k 
$B^\pm \rightarrow D^0 K^\pm$ and $\sim$2k $B^\pm \rightarrow \bar{D^0} K^\pm$ 
decays are expected in one year, and the sensitivity to $\gamma$
is $\sigma(\gamma)\sim 5^\circ$ considering the background ($\sim 3.9^\circ$
with no background, see Fig. \ref{dsk} (right) for an example of
the fit results). The sensitivity with lower values of $r_B$ is under study.

%with the neutral D decaying into $K \pi$
%and $K \pi \pi \pi$, are measured 
% The expected precision is about $5^\circ$ after
%1 year of data taking.

\begin{figure}[!t]
  \includegraphics[height=.15\textheight]{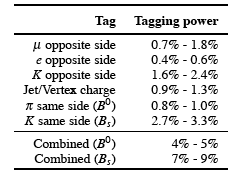}
  \hspace{10mm}
  \includegraphics[height=.10\textheight]{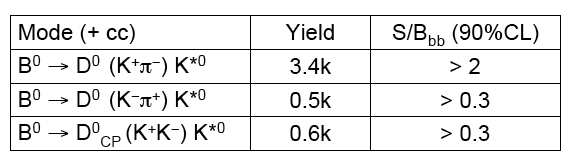}
  \caption{Tagging power (left).
Annual yield and $S/B$ ratio for the decays $B_d \rightarrow D^0 K ^{*0}$
(right).}
  \label{table}
\end{figure}

The decay $B_d \rightarrow D^0 K^{*0}$ also proceeds via a tree diagram,
and has no sensitivity to new physics. The method \cite{glw} 
to extract $\gamma$
is based on the measurement of six time-integrated decay rates
for $B_d \rightarrow D^0 K^{*0}$, $B_d \rightarrow \bar{D^0} K^{*0}$,
$B_d \rightarrow D^0_{CP} K^{*0}$ and their CP conjugates.
Samples of about 3.4k $B_d \rightarrow \bar{D^0}(K^- \pi^+)K^{*0}$,
0.5k $B_d \rightarrow D^0(K^+ \pi^-)K^{*0}$, and 
0.6k $B_d \rightarrow D^0_{CP}(K^+ K^-)K^{*0}$ decays will 
be reconstructed per year (see Table \ref{table} (right)). 
The precision reached is 
$\sigma(\gamma) \sim 8^\circ$ with 2$fb^{-1}$.

$\gamma$ can be measured from time-dependent asymmetries
for $B_d \rightarrow \pi \pi$ and $B_s \rightarrow KK$ decays 
\cite{battaglia,fleischer1,fleischer2}, where
both penguin and tree diagrams contribute.
This measurement is sensitive to new physics appearing in the penguin loops.
Here again the $K/\pi$ separation is crucial because of numerous
reflections from $B_{d,s}$ and $\Lambda_b$ decays.
The asymmetries can be parameterised as $A_{CP} = A^{dir} cos(\Delta mt) +
A^{mix} sin(\Delta mt)$. The four observables are functions of the
parameters: $\gamma$, mixing phases $\phi_d$ and $\phi_s$, 
ratios of penguin over tree contributions $P/T = d e^{i\theta}$. 
Measuring $\phi_d$ and $\phi_s$
independently and assuming U-spin symmetry ($d_{\pi\pi} = d_{KK}$,
$\theta_{\pi\pi} = \theta_{KK}$), the relations above reduce
to four measured quantities with three unknowns, which can be solved
for $\gamma$. The annual yield of 26k $B_d \rightarrow \pi \pi$ ($B/S < 0.7$)
and 37k $B_s \rightarrow KK$ ($B/S < 0.3$) 
events allows a precision of $\sigma(\gamma) \sim 5^\circ$.
Since the system is overconstrained, the hypothesis of U-spin conservation
can be also tested.

%\end{itemize}

\subsection{Rare decays: $B \rightarrow K^* \mu \mu$}

Rare loop-induced decays are sensitive to new physics in many Standard Model
extensions. At LHCb, rare decays such as decay $B_d \rightarrow K^* \mu \mu$
will be studied. The expected annual yield is $\sim$4.5k, 
with a ratio $B/S < 0.2$,
using the Standard Model prediction for the branching ratio of $\sim 10^{-6}$. 
Further improvements are expected in the signal selection to enhance
the yield.

\begin{figure}[!t]
  \includegraphics[height=.24\textheight]{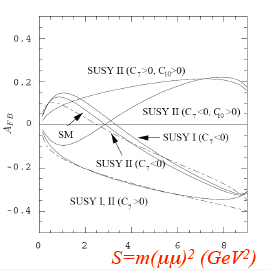}
  \hspace{10mm}
  \includegraphics[height=.24\textheight]{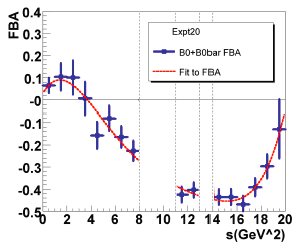}
  \caption{Distribution of the forward-backward asymmetry in 
$B_d \rightarrow K^* \mu \mu$, as a function of the $\mu\mu$ invariant mass,
for different extensions of the Standard Model (left).
Distribution of the forward-backward asymmetry as expected in LHCb
with $10 fb^{-1}$ (right).}
  \label{kstmumu}
\end{figure}

This channel is well suited to searches for
new physics, since New Physics models make definite predictions for
the shape of the forward-backward asymmetry of the $\mu^+$ in the
$\mu\mu$ rest frame
with respect to the $B$ direction, as a function of the
$\mu\mu$ invariant mass \cite{rare}
(see Fig. \ref{kstmumu} (left)). In particular the zero crossing of
the forward-backward asymmetry is predicted 
with small theoretical uncertainties.
With a year of data taking, a clear observation
of NP should be possible; with $10 \ fb^{-1}$ the position of the
zero should be located with a precision of $\pm 0.53$ GeV$^2$
(see Fig \ref{kstmumu} (right)).

\section{Conclusions}
Using a detailed simulation of the LHCb detector, we demonstrate that
the LHCb experiment can efficiently reconstruct many different 
decay modes with a very good performance in the trigger, 
proper time resolution, mass resolution, and flavour tagging.
This will allow LHCb to fully explore the $B_s$ mixing, to extract
CKM parameters with various methods, and
to perform studies of rare decays. The experiment will make 
precision tests of the Standard Model in order to constrain,
and possibly discover, new physics.

%%%%%%%%%%%%%%%%%%%%%%%%%%%%%%%%%%%%%%%%%%%%%%%%
%% BACKMATTER
%%%%%%%%%%%%%%%%%%%%%%%%%%%%%%%%%%%%%%%%%%%%%%%%

\begin{theacknowledgments}
The author would like to thank
I.~Belyaev,
O.~Deschamps,
H.~Dijkstra,
R.~Forty,
C.~R.~Jones,
T.~Ruf,
O.~Schneider
G.~Wilkinson,
and Y.~Xie
for their help in preparing these proceedings.
\end{theacknowledgments}

%%%%%%%%%%%%%%%%%%%%%%%%%%%%%%%%%%%%%%%%%%%%%%%%
%% References
%%%%%%%%%%%%%%%%%%%%%%%%%%%%%%%%%%%%%%%%%%%%%%%%

\bibliographystyle{aipproc}   % if natbib is available
\bibliography{LHCb_phys}

\hyphenation{Post-Script Sprin-ger}
\begin{thebibliography}{20}
\expandafter\ifx\csname natexlab\endcsname\relax\def\natexlab#1{#1}\fi
\providecommand{\enquote}[1]{``#1''}
\expandafter\ifx\csname url\endcsname\relax
  \def\url#1{\texttt{#1}}\fi
\expandafter\ifx\csname urlprefix\endcsname\relax\def\urlprefix{URL }\fi
\providecommand{\eprint}[2][]{\url{#2}}

\bibitem[LHCb(2001{\natexlab{a}})]{VeloTDR}
LHCb, {VELO Technical Design Report}, Tech. rep., CERN/LHCC/2001-011
  (2001{\natexlab{a}}).

\bibitem[LHCb(2002)]{ITTDR}
LHCb, {Inner Tracker Technical Design Report}, Tech. rep., CERN/LHCC/2002-029
  (2002).

\bibitem[LHCb(2001{\natexlab{b}})]{OTTDR}
LHCb, {Outer Tracker Technical Design Report}, Tech. rep., CERN/LHCC/2001-024
  (2001{\natexlab{b}}).

\bibitem[LHCb(2000{\natexlab{a}})]{RichTDR}
LHCb, {RICH Technical Design Report}, Tech. rep., CERN/LHCC/2000-037
  (2000{\natexlab{a}}).

\bibitem[LHCb(2000{\natexlab{b}})]{CaloTDR}
LHCb, {CALO Technical Design Report}, Tech. rep., CERN/LHCC/2000-036
  (2000{\natexlab{b}}).

\bibitem[LHCb(2001{\natexlab{c}})]{MuonTDR}
LHCb, {MUON Technical Design Report}, Tech. rep., CERN/LHCC/2001-010
  (2001{\natexlab{c}}).

\bibitem[Pietrzyk(2006)]{Bolek}
B.~Pietrzyk, \enquote{{LHCb Detector Status and Commissioning},} in
  \cite{hcp2006}.

\bibitem[Jones(2006)]{Jonesc}
C.~R. Jones, \enquote{{Tracking, Vertexing and Particle Identification in
  LHCb},} in  \cite{hcp2006}.

\bibitem[LHCb(2003)]{ReopTDR}
LHCb, {Reoptimised Dector Design and Performance Technical Design Report},
  Tech. rep., CERN/LHCC/2003-030 (2003).

\bibitem[{CDF Collaboration}(2006)]{CDF}
{CDF Collaboration}, \emph{{Physics Review Letters}} \textbf{97}, 062003
  (2006).

\bibitem[{D0 Collaboration}(2006)]{D0}
{D0 Collaboration}, \emph{{Physics Review Letters}} \textbf{97}, 021802 (2006).

\bibitem[{A. Snyder and H.R. Quinn}(1993)]{alpha}
{A. Snyder and H.R. Quinn}, \emph{{Physics Review D}} \textbf{48}, 2139 (1993).

\bibitem[{D. Atwood, I. Dunietz, A. Soni}(1997)]{ads}
{D. Atwood, I. Dunietz, A. Soni}, \emph{{Physics Review Letters}} \textbf{78},
  3275 (1997).

\bibitem[{D. Atwood, I. Dunietz, A. Soni}(2001)]{ads1}
{D. Atwood, I. Dunietz, A. Soni}, \emph{{Physics Review D}} \textbf{63}, 036005
  (2001).

\bibitem[{M. Gronau and D. Wyler}(1991)]{glw}
{M. Gronau and D. Wyler}, \emph{{Physics Letters B}} \textbf{265}, 172 (1991).

\bibitem[{M. Battaglia, A.J. Buras, P. Gambino, A. Stocchi}(2003)]{battaglia}
{M. Battaglia, A.J. Buras, P. Gambino, A. Stocchi}, \emph{CERN-TH-2003-002}
  (2003).

\bibitem[Fleischer(1999)]{fleischer1}
R.~Fleischer, \emph{{Physics Letters B}} \textbf{459}, 306 (1999).

\bibitem[{R. Fleischer, J. Matias}(2002)]{fleischer2}
{R. Fleischer, J. Matias}, \emph{{Physics Review D}} \textbf{66}, 054009
  (2002).

\bibitem[{T. Goto, Y. Okada, Y. Shimizu, M. Tanaka}(1997)]{rare}
{T. Goto, Y. Okada, Y. Shimizu, M. Tanaka}, \emph{{Physics Review D}}
  \textbf{55}, 4273 (1997).

\bibitem[hcp(2006)]{hcp2006}
\emph{HCP 2006 Proceedings}, 2006.

\end{thebibliography}

\end{document}